\documentclass[a4paper]{jpconf}
\usepackage{graphicx}
\begin{document}
\title{Beam spin asymmetry in deeply virtual $\pi$ production}

\author{Murat M. Kaskulov and Ulrich Mosel}

\address{Institut f\"ur Theoretische Physik, Universit\"at Giessen,
             D-35392 Giessen, Germany}

\ead{murat.kaskulov@theo.physik.uni-giessen.de}

\begin{abstract}
An interpretation of the beam spin azimuthal asymmetries measured at JLAB
in deep exclusive electroproduction of charged and neutral pions is
presented. The model combines a Regge pole approach with the effect of
nucleon resonances.
The $s$- and $u$-channel contributions are described using a dual Bloom-Gilman
connection between the exclusive form 
factors and inclusive deep inelastic structure functions. The results are in
agreement with data provided the excitations of nucleon resonances are
taken into account. 
\end{abstract}

\section{Introduction} 
Electroproduction of mesons in the deep inelastic scattering (DIS),
$\sqrt{s}>2$~GeV and $Q^2>1$~GeV$^2$, 
is a modern tool which permits to study the structure of the nucleon on the partonic
level. Exclusive channels in DIS are of particular importance. 
In this kind of hard processes one may learn about the off-forward 
parton distributions  that parameterize an intrinsic
nonperturbative pattern of the nucleon, see Ref.~\cite{Weiss:2009ar} and references
therein. Much work have been done to 
understand the production of pions in exclusive kinematics.
For instance, in QCD at large values of $(\sqrt{s},Q^2)$
and finite value of Bjorken $x_{\rm B}$  the description of $N(e,e'\pi)N'$ 
relies on the dominance of the longitudinal cross 
section $\sigma_{\rm L}$~\cite{Collins:1996fb}.  The transverse part
$\sigma_{\rm T}$ is
predicted to be  suppressed by power of $\sim 1/Q^2$. 
However, being a leading twist prediction the kinematic domain where this
power suppresion dominates is not
yet known for $\pi$ production. 
A somewhat different concept is used in Regge pole
models which rely on effective hadronic
degrees of freedom. Here
the exclusive $(\gamma^*,\pi)$ forward production mechanism is
peripheral, 
that is a sum of all possible
$t$-channel meson-exchange processes.
Although both partonic and Regge descriptions are presumably dual 
the exclusive reactions
have a potential to discriminate between different models.
Related studies have been carried out at
JLAB~\cite{Horn:2007ug,Collaboration:2010kna} and at DESY~\cite{:2007an}.   
A dedicated program on
exclusive production of pions is 
planned in the future at the JLAB upgrade.

On the experimental
side, it is tempting to see an onset of $\sigma_{\rm L}/\sigma_{\rm T} \propto Q^2$ scaling
at presently available energies.
However, the high $Q^2$ data from
JLAB and single spin asymmetries measured in true DIS events at HERMES~\cite{:2009ua}
show clearly nonvanishing
transverse components in $p(\gamma^*,\pi^+)n$.
At JLAB~\cite{Horn:2007ug,:2009ub,Tadevosyan:2007yd,Blok:2008jy}, 
DESY~\cite{Desy,Ackermann:1977rp,Brauel:1979zk}, Cornell~\cite{Cornell_1,Cornell_2,Cornell_3}
and CEA~\cite{{CEA}} the high $Q^2$ region  is dominated by the conversion of
transverse photons in
$\sigma_{\rm T}$.  
For instance, the $Q^2$ dependence of the partial $\sigma_{\rm L}$ and $\sigma_{\rm T}$
cross sections in the
$\pi^{+}$ electroproduction above $\sqrt{s}>2$~GeV  has been studied
in~\cite{Horn:2007ug}. In the
charged pion case  the 
longitudinal cross section $\sigma_{\rm L}$ at forward angles is well
described by the quasi-elastic  $\pi$ knockout
mechanism~\cite{Sullivan:1970yq,Neudatchin:2004pu}. It is driven by
the pion charge form factor~\cite{Horn:2006tm,Huber:2008id} both at JLAB and HERMES.
On the contrary, the $(\sqrt{s},Q^2)$ behavior of $\sigma_{\rm T}$ remains to
be puzzling. The data demonstrate that $\sigma_{\rm T}$ is large and tends to increase relative to
$\sigma_{\rm L}$ as a function of $Q^2$. Interestingly, the $(\sqrt{s},Q^2)$
dependence of exclusive $\sigma_{\rm T}$ exhibits features similar to that in
$p(e,e'\pi^{+})X$ semi-inclusive  
 cross sections in DIS in the limit $z\to
1$~\cite{Kaskulov:2008xc,Kaskulov:2009gp}. This kind of an exlusive-inclusive connection~\cite{Bjorken:1973gc} has been also
observed in exclusive $(\gamma^*,\rho^0)$ production~\cite{Gallmeister:2010wn}.
On the theoretical side, hadronic models based on the meson-exchange scenario
alone largely
underestimate the measured $\sigma_{\rm T}$ in electroproduction, see
Ref.~\cite{Blok:2008jy} for further discussions and references therein. 

\begin{figure}[t]
\begin{center}
\includegraphics[clip=true,width=1\columnwidth,angle=0.]
{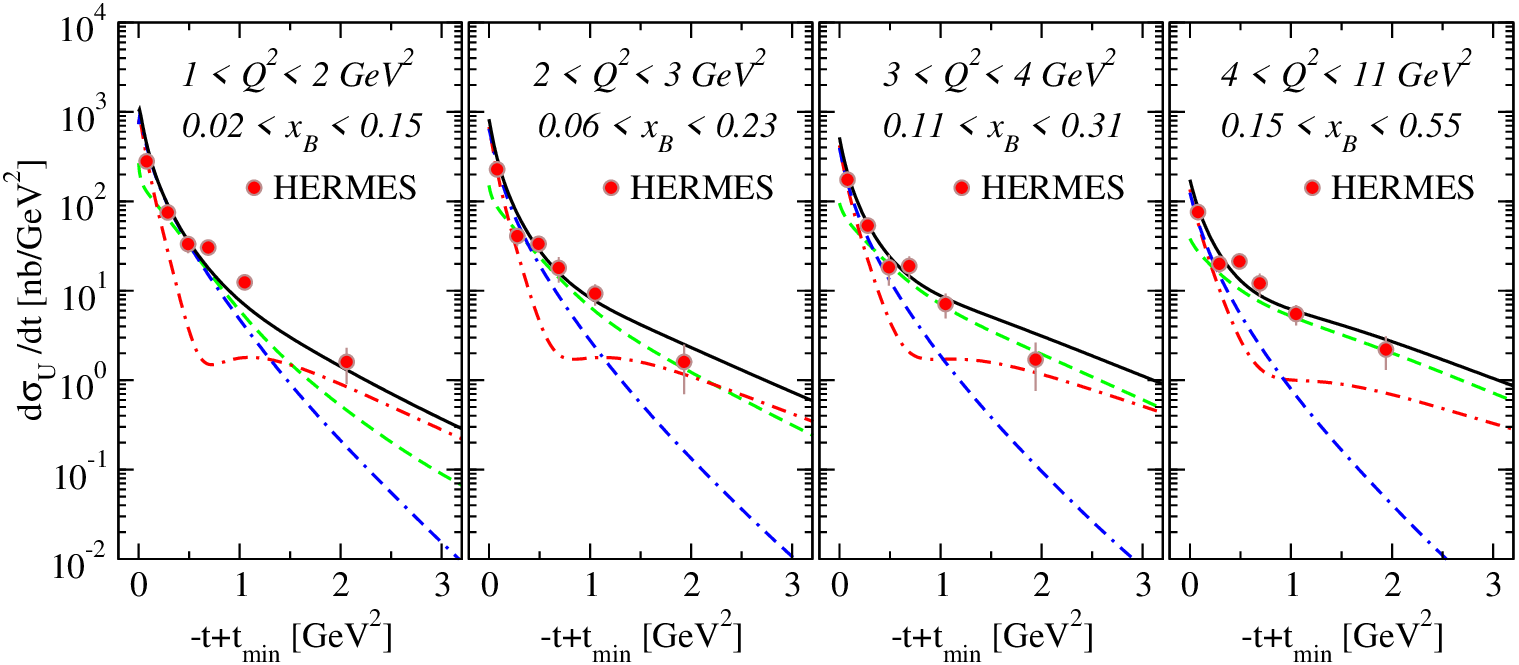}
\begin{minipage}[b]{38pc} \caption{\label{EffHermes} \small $-t+t_{min}$ dependence of the 
differential cross section $d\sigma_U/dt = d\sigma_{\rm T}/dt + \epsilon d\sigma_{\rm
L}/dt$ in exclusive reaction $p(\gamma^*,\pi^+)n$ at HERMES. 
The experimental data are from Ref.~\cite{:2007an}. 
The calculations are performed for the average values of
$(Q^2,x_{\rm B})$ in a given $Q^2$ and Bjorken $x_{\rm B}$ bin.
The solid curves are the full model results.
The dash-dotted curves correspond to the longitudinal $\epsilon d\sigma_{\rm L}/dt$
and the dashed curves to the transverse $d\sigma_{\rm T}/dt$ components of the
cross section.
The dash-dash-dotted curves describe the results without the
resonance/partonic effects.}
\vspace{-0.5cm}
\end{minipage}
\end{center}
\end{figure}

Phenomenological solutions of the $\sigma_{\rm T}$ problem already
exist in the literature~\cite{Kaskulov:2008xc,KM}. The
description of charged pion production proposed in Ref.~\cite{KM} relies on
the residual contribution of the nucleon resonances.  It is supposed that the excitations of nucleon resonances dominate
in electroproduction. The resonances are dual to the direct partonic
interactions due to the Bloom-Gilman duality connection and, correspondingly,
their form factors are determined by parton distribution functions. The $s(u)$-channel resonances
supplement the Reggeon exchanges in the $t$-channel. Therefore, we distinguish
peripheral $t$-channel meson-exchange processes and the $s(u)$-channel resonance/partonic
contributions. In this way all the data collected so far in the
charged pion electoproduction $(e,e'\pi^{\pm})$ at JLAB, DESY, Cornell and CEA
can be well described~\cite{KM}. As an example, in Fig.~\ref{EffHermes} we show our results
for the $-t+t_{min}$ dependence of the 
differential cross section $d\sigma_U/dt = d\sigma_{\rm T}/dt + \epsilon d\sigma_{\rm
L}/dt$ in exclusive reaction $p(\gamma^*,\pi^+)n$ at HERMES.

\section{Beam single spin asymmetry (SSA)}
In this talk we consider the electroproduction reaction
\begin{equation}
\vec{e} + N \to e' + \pi + N
\end{equation}
Here we assume that the target nucleon is unpolarized, whereas we allow
arbitrary polarization for the incoming electron.
With a polarized beam 
$\vec{e}$ and with an unpolarized target there is an additional 
component $\sigma_{\rm LT'}$~\cite{KM} in the $(e,e'\pi)$ cross section which is
proportional to the imaginary part of an interference between the \textsc{l/t}
photons and therefore sensitive to the relative phases of amplitudes.  

Using the polarized electron beam, the longitudinal beam single-spin asymmetry (SSA) in $N(\vec{e},e'\pi)N'$ scattering is defined
so that
\begin{equation}
\label{BSSA}
A_{\rm LU}(\phi) \equiv
\frac{d\sigma^{\rightarrow}(\phi)-d\sigma^{\leftarrow}(\phi)}{d\sigma^{\rightarrow}(\phi)+d\sigma^{\leftarrow}(\phi)},
\end{equation}
where $d\sigma^{\rightarrow}$ refers to positive helicity $h=+1$ of the incoming
electron. The azimuthal moment associated with the beam SSA is
given by
\begin{equation}
\label{BeamSSAmoment}
A^{\sin(\phi)}_{\rm LU} = \frac{\sqrt{2\varepsilon(1-\varepsilon)}d\sigma_{\rm
    LT'}}{d\sigma_{\rm T} + \varepsilon d\sigma_{\rm L}}.
\end{equation}

In general, a nonzero $\sigma_{\rm LT'}$ or the corresponding 
beam SSA $A_{\rm LU}(\phi)$, Eq.~(\ref{BSSA}), demands 
interference between single helicity flip and nonflip or double helicity flip 
amplitudes. In Regge models the asymmetry may result from 
Regge cut corrections to single reggeon exchange. 
This way the amplitudes in the product acquire different phases and therefore 
relative imaginary parts. A nonzero beam SSA  can be also generated by the 
interference pattern of amplitudes where particles with opposite
parities are exchanged.

\section{Beam SSA in charged pion production}
\begin{figure}[h]
\includegraphics[clip=true,width=0.6 \columnwidth,angle=0.]{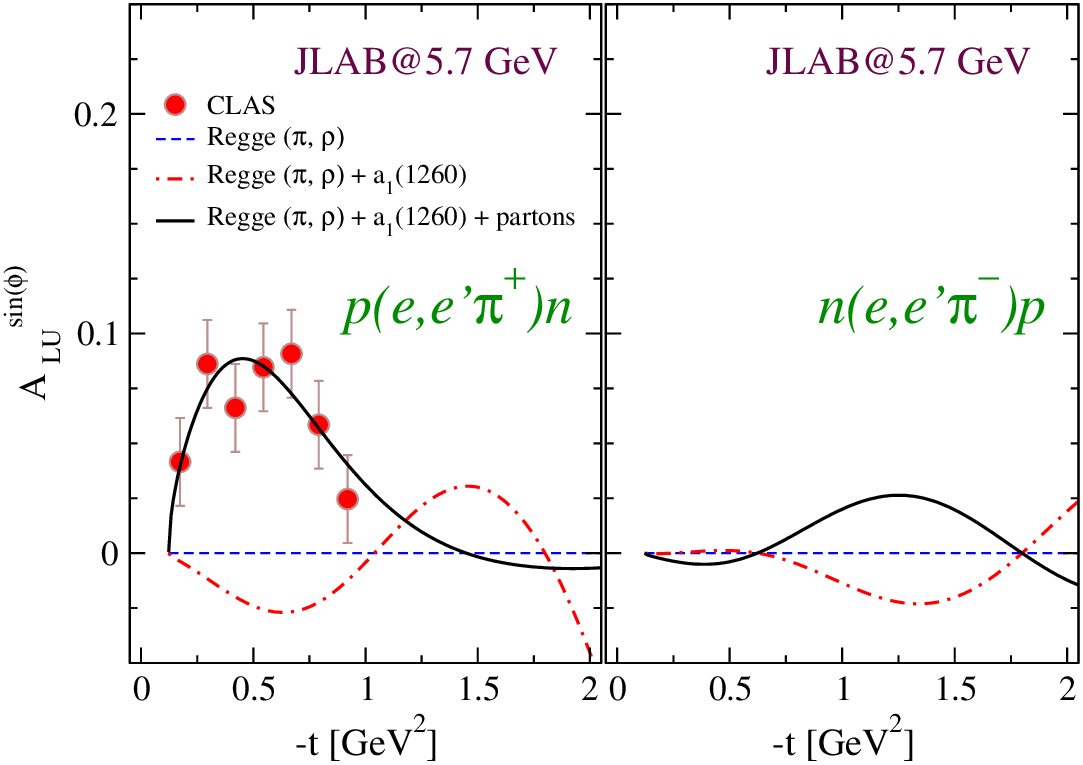}
\begin{minipage}[b]{15pc} \caption{\label{BSAAvakian}
\small  Left panel: The beam spin azimuthal moment
$A^{\sin(\phi)}_{\rm LU}$ in exclusive
reaction $p(\gamma^*,\pi^+)n$ as a function of $-t$. 
The CLAS/JLAB data are from~\cite{Avakian:2004dt}.  
The dashed curves describe 
the results (the asymmetry is zero) without the resonance contributions 
and neglecting the exchange of unnatural parity  $a_1(1260)$ Regge trajectory.  
The dash-dotted curves correspond
to the addition of the axial-vector $a_1(1260)$-reggeon exchange. 
The solid curves are the model results and account for the
resonance/partonic effects.
Right panel:
The beam spin azimuthal moment $A^{\sin(\phi)}_{\rm LU}$ in exclusive reaction
$n(\gamma^*,\pi^-)p$.
\vspace{-0.0cm}
}
\end{minipage}
\end{figure}

In the left panel of Fig.~(\ref{BSAAvakian}) we plot the CLAS 
data~\cite{Avakian:2004dt} for the azimuthal moment $A^{\sin(\phi)}_{\rm LU}$ 
associated with the beam SSA, Eq.~(\ref{BeamSSAmoment}), in the reaction 
$p(\vec{e},e'\pi^+)n$.  These data have been collected in hard scattering
kinematics $E_e=5.77$~GeV, $W>2$~GeV and $Q^2>1.5$~GeV$^2$. 
The experiment shows a sizable and positive beam SSA. 

In the left and right panels of Fig.~(\ref{BSAAvakian}) 
we present our results for the azimuthal moments $A^{\sin(\phi)}_{\rm LU}$ 
in the reactions $p(\vec{e},e'\pi^+)n$ and $n(\vec{e},e'\pi^-)p$,
respectively. 

At first, we consider $A^{\sin(\phi)}_{\rm LU}$ generated by the exchange of 
Regge trajectories. In Fig.~(\ref{BSAAvakian}) the dashed curves describe  
the model results without the effects of resonances and neglecting the exchange of the
axial-vector $a_1(1260)$ Regge trajectory. This model results in a zero
$A^{\sin(\phi)}_{\rm LU}$ and therefore a zero beam SSA. The addition of the unnatural
parity $a_1(1260)$-exchange generates by the interference with the natural parity 
$\rho(770)$ exchange a sizable $A^{\sin(\phi)}_{\rm LU}$ in both channels. This result
corresponds to the dash-dotted curves in Fig.~(\ref{BSAAvakian}). 
In the rest of unpolarized observables the effect of the axial-vector 
$a_1(1260)$ is small. However, as one can see, the contribution of $a_1(1260)$ 
is important in the polarization observables. For instance, a strong
interference pattern of the $a_1(1260)$-reggeon exchange makes the
polarization  observables, like the beam SSA,  very sensitive to the different 
scenarios describing the structure and behavior 
of $a_1(1260)$ in high-$Q^2$ processes.
In the last step we account for the resonance contributions. The latter
strongly influence the asymmetry parameter
$A^{\sin(\phi)}_{\rm LU}$. The model results (solid curves) are in agreement 
with the positive $A^{\sin(\phi)}_{\rm LU}$ in the
$\pi^+$ channel and predict much smaller $A^{\sin(\phi)}_{\rm LU}$ in the $\pi^-$
channel.  A sizable and positive $A^{\sin(\phi)}_{\rm LU}$ has been also observed at HERMES
in $\pi^+$ SIDIS close to the exclusive limit $z\to
1$~\cite{Airapetian:2006rx}. 

\begin{figure}[h]
\begin{center}
\begin{minipage}[b]{11pc} \caption{\label{BeamSSApi0} 
\small The beam spin azimuthal moment
$A^{\sin(\phi)}_{\rm LU}$ in exclusive
reaction $p(\gamma^*,\pi^0)p$ as a function of $-t$ for different $(Q^2,x_{\rm
B})$ bins.
The solid curves are the model results and account for the
residual effect of nucleon resonances. The experimental data are from~\cite{DeMasi:2007id}.
}
\end{minipage}
\includegraphics[clip=true,width=0.7\columnwidth,angle=0.]
{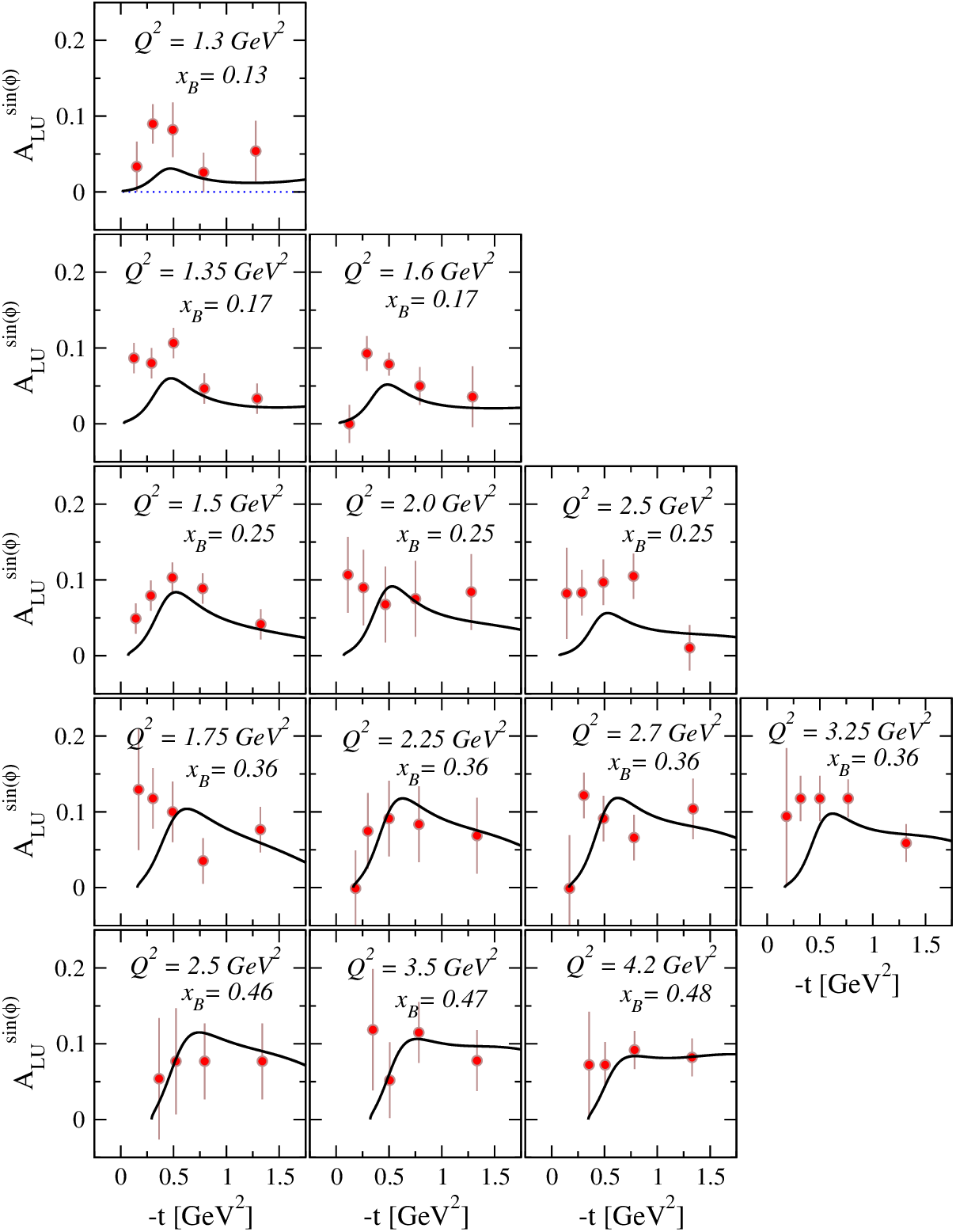}
\vspace{-0.5cm}
\end{center}
\end{figure}

\section{Beam SSA in neutral pion production}
As in deep exclusive $\pi^+$ electroproduction a sizeable and positive beam SSA   
has recently been  measured at CLAS/JLAB also in the exclusive reaction $p(\vec{e},e'\pi^0)p$~\cite{DeMasi:2007id}. 
It was shown that the simple Regge model used in~\cite{DeMasi:2007id} fails 
to explain the measured kinematic $(\sqrt{s},Q^2)$  dependencies. 
We have, therefore, extended our calculation of Ref.~\cite{KM} to the neutral pion channel. In the
Regge exchange contributions the vector $\omega(782)$, $\rho(770)$ and axial-vector
$b_1(1235)$ and $h_1(1170)$ trajectories are taken into account. We find that at
high values of $Q^2$ the dominant contribution to the beam SSA again comes from
the residual excitation of nucleon resonances. Our results are shown in Fig.~\ref{BeamSSApi0} and
describe the JLAB data very well.

\section{Summary}
In summary, a description of exclusive pion electroproduction 
$(e,e'\pi)$ off nucleons at high energies is proposed.
Following the two-component hadron-parton picture of Refs.~\cite{Kaskulov:2008xc,Kaskulov:2009gp}
the  model of Ref.~\cite{KM} combines a Regge pole approach with residual effects of nucleon
resonances. The contribution of nucleon
resonances has been assumed to be dual to direct partonic
interaction and therefore describes the hard part of the model cross
sections.

In this talk we presented the results for the cross sections with
longitudinally polarized electron beam. 
We have shown that the resonance/partonic mechanism
is responsible for the positive azimuthal beam  SSA  observed in
exclusive reactions $p(\vec{e},e'\pi^+)n$ and $p(\vec{e},e'\pi^0)p$.
On the contrary, the beam SSA in deep exclusive $\pi^-$ production off the neutrons
is predicted to be much smaller in magnitude and very sensitive to
the different scenarios concerning the structure of the $a_1(1260)$
axial-vector meson.

\ack
This work was supported by DFG through TR16 and by BMBF.

\end{document}